\documentclass[a4paper,fleqn]{cas-db}

\usepackage[numbers]{natbib}
\usepackage{enumitem}
\usepackage{color}
\usepackage{mathtools}
\usepackage{amsmath}
\usepackage{graphicx}
\usepackage{tcolorbox}
\usepackage{xcolor}
\definecolor{Ao}{rgb}{0.0, 0.5, 0.0}
\def\tsc#1{\csdef{#1}{\textsc{\lowercase{#1}}\xspace}}
\tsc{WGM}
\tsc{QE}

\definecolor{janekgreen}{rgb}{0.0, 0.5, 0.25}

\begin{document}
\let\WriteBookmarks\relax
\def\floatpagepagefraction{1}
\def\textpagefraction{.001}
% \thispagestyle{fancy}
% \ifthenelse{\boolean{shortarticle}}{\ifthenelse{\boolean{singlecolumn}}{\abscontentformatted}{\abscontent}}{} 
\shorttitle{Semantic segmentation of multispectral photoacoustic images using deep learning}
% Short author
\shortauthors{Schellenberg et al.}  

% Main title of the paper
\title[mode = title]{Semantic segmentation of multispectral photoacoustic images using deep learning}

\author[1,2,3]{Melanie Schellenberg}[]\cormark[1]
\author[1,4] {Kris K. Dreher}
\author[1] {Niklas Holzwarth}
\author[5] {Fabian Isensee}
\author[1,2,5] {Annika Reinke}
\author[6]{{Nicholas Schreck}}
\author[1] {Alexander Seitel}
\author[1] {Minu D. Tizabi}
\author[1,2,3,5,7] {Lena Maier-Hein}\fnmark[1]\cormark[1]
\author[1]{Janek Gröhl}[]\fnmark[1,2]

\address[1]{Computer Assisted Medical Interventions (CAMI), German Cancer Research Center (DKFZ), Heidelberg, Germany}
\address[2]{Faculty of Mathematics and Computer Science, Heidelberg University, Heidelberg, Germany}
\address[3]{HIDSS4Health - Helmholtz Information and Data Science School for Health, Heidelberg, Germany}
\address[4]{Faculty of Physics and Astronomy, Heidelberg University, Heidelberg, Germany}
\address[5]{HI Applied Computer Vision Lab, Division of Medical Image Computing, German Cancer Research Center (DKFZ), Heidelberg, Germany}
\address[6]{{Division of Biostatistics, German Cancer Research Center (DKFZ), Heidelberg, Germany}}
\address[7]{Medical Faculty, Heidelberg University, Heidelberg, Germany}

\fntext[1] {Shared last authorship.}
\cortext[1]{Send correspondence to
melanie.schellenberg@dkfz-heidelberg.de (M. Schellenberg) or l.maier-hein@dkfz-heidelberg.de (L. Maier-Hein).} 
\fntext[2]{Now at Cancer Research UK Cambridge Institute, University of Cambridge, Robinson Way, Cambridge, CB2 0RE, U.K.} 

\begin{abstract}
Photoacoustic (PA) imaging has the potential to revolutionize functional medical imaging in healthcare due to the valuable information on tissue physiology contained in multispectral photoacoustic measurements. Clinical translation of the technology requires conversion of the high-dimensional acquired data into clinically relevant and interpretable information. In this work, we present a deep learning-based approach to semantic segmentation of multispectral photoacoustic images to facilitate image interpretability. Manually annotated photoacoustic {and ultrasound} imaging data are used as reference and enable the training of a deep learning-based segmentation algorithm in a supervised manner. Based on a validation study with experimentally acquired data from 16 healthy human volunteers, we show that automatic tissue segmentation can be used to create powerful analyses and visualizations of multispectral photoacoustic images. Due to the intuitive representation of high-dimensional information, such a preprocessing algorithm could be a valuable means to facilitate the clinical translation of photoacoustic imaging.
\end{abstract}

\begin{keywords}{Medical Image Segmentation \sep Deep Learning \sep Multispectral Imaging \sep Photoacoustics \sep Optoacoustics}
\end{keywords}

\maketitle

\section{Introduction}
\begin{figure}[h!tb]
    \centering
    {
    \includegraphics[width=\columnwidth]{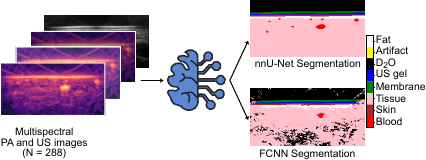}}
    \caption{\textbf{Overview of the proposed approach to automatic semantic image annotation.} {The nnU-Net} and {a} fully-connected neural network (FCNN) automatically create semantic annotations of multiple tissue types, including skin, blood, and fat, based on multispectral photoacoustic (PA) and  ultrasound (US) images. }
    \label{fig:overview}
\end{figure}

Photoacoustic (PA) imaging (PAI) is an emerging and rapidly developing imaging modality that enables real-time, non-invasive, and radiation-free measurement of optical tissue properties~\cite{xia2014photoacoustic}. PAI has the potential to spatially resolve valuable morphological and functional tissue information, such as the blood oxygen saturation (sO$_2$)~\cite{brunker2017photoacoustic}, for up to several centimeters in depth~\cite{beard2011biomedical}. While the recovery of accurate and reliable functional parameters from PA measurements is an ongoing field of research {~\cite{hauptmann2020deep, liu2021five, triki2021h}}, providing accurate and interpretable visualizations of multispectral PA measurement data is a crucial step towards the clinical translation of PAI. One way to achieve this could be to classify tissue pixel-wise based on the multispectral PA signal, thus segmenting the tissue into disjunct regions, as illustrated in Figure ~\ref{fig:overview}. These can be annotated with relevant information, such as structure-specific sO$_2$.\\

Several groups have already worked on methods for automatic image segmentation in PAI, for example for the automatic identification of structures in small animal images~\cite{lafci2020deep, lafci2020efficient, Liang2021}, the segmentation of breast cancer~\cite{zhang2018photoacoustic}, or for vessel segmentation both in simulation studies~\cite{luke2019net} and experimental settings~\cite{chlis2020sparse,yuan2020hybrid}. Furthermore, work has been conducted towards the annotation of different skin layers in raster-scanned images~\cite{gerl2020distance, Ly2021, moustakidis2019fully}. However, to our knowledge, no work has been published to date on the automatic \emph{multi-label} semantic annotation of multispectral PA images in humans.

The purpose of this paper is, therefore, to address this gap in the literature. Specifically, we investigate the hypothesis that \textit{automatic \textit{multi-label} PA image segmentation with neural networks is feasible.} More specifically, we explore two different approaches to utilizing the multispectral data as the input of a neural network: 1) a \textit{single-pixel} representation for a fully-connected neural network (FCNN) and 2) a \textit{full image} representation for {the nnU-Net~\cite{isensee2021nnu}}. As many commonly used PA devices capture PA and ultrasound (US) data simultaneously~\cite{becker2018multispectral,wei2015real}, we additionally compare the performance of our method for different types of input images, namely PA images, US images and combined PA and US images which we refer to as PAUS images. Our methods are {trained, }validated{, and tested }on {separate splits of} a data set of forearm, calf, and neck measurements acquired from 16 healthy volunteers.

\section{Materials and methods}

In this section, we describe the data set we acquired and annotated for the training and validation of our method (cf. Section~\ref{sec:meth:data}) and the deep learning-based methods for semantic annotation of PA data (cf. Section~\ref{sec:meth:dl}). 

\subsection{Data}
\label{sec:meth:data}
The following section provides details of the acquisition (cf. Section \ref{sec:meth:data_acq}) and annotation (cf. Section{s} \ref{sec:meth:data_anno} {and \ref{sec:meth:IRV}}) of the PA and US data as well as specifics of the data split (cf. Section \ref{sec:meth:data_split}).  
\subsubsection{Data acquisition}
\label{sec:meth:data_acq}

\begin{figure}[h!tb]
\centering
\includegraphics[width=\columnwidth]{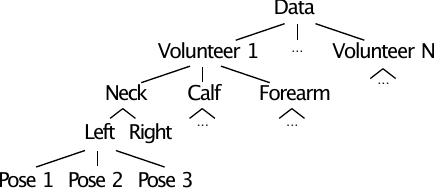}
\caption{\textbf{Visualization of the hierarchical nature of the data.} For each volunteer, the neck, calf, and forearm were imaged at three distinct poses both on the left and right side of the body.}
\label{fig:hierarchy}
\end{figure}

The data set consisted of multispectral PA and US images of 16 healthy human volunteers. For each of the 16 volunteers, the forearm, calf, and neck were imaged at three distinct locations both on the left and right side of the body (cf. Figure ~\ref{fig:hierarchy} for the hierarchical representation of the data), yielding N\,=\,288 unique multispectral PA and US image pairs in total. {These locations were chosen as they are easily accessible, the least intrusive, and feature superficial blood vessels, which were of key interest for this study. 18 scans were acquired per volunteer as a trade-off between acquisition time ($\sim$1h) and number of scans.} Ethics approval was obtained from the committee of the medical faculty of Heidelberg University under reference number S-451/2020 and the study was registered with the German Clinical Trials Register under reference number DRKS00023205. The experiments were carried out in accordance with relevant guidelines and regulations of the ethics approval, as for example laser safety guidelines, and informed consent was obtained from all subjects.

The images were acquired using the multi-spectral optoacoustic tomography (MSOT) Acuity Echo device (iThera Medical, Munich, Germany) using 26 wavelengths equidistantly chosen between 700\,nm and 950\,nm{, similar to ~\cite{chlis2020sparse} and allowing the results of this study to be easily applied to data-driven oximetry~\cite{grohl2021learned}}. Each location was imaged freehand and as statically as possible for approximately 30 seconds. The US images were reconstructed using a proprietary backprojection algorithm provided by the vendor and the PA images were reconstructed using a custom implementation of the backprojection algorithm~\cite{kirchner2018signed} within the Medical Imaging Interaction Toolkit (MITK)~\cite{nolden2013medical}. The sequences were post-processed including a correction for variations in the laser pulse energy and an optimization of the Signal-to-Noise-Ratio (SNR). To correct for the variations in laser pulse energy, the PA images were divided by the respective laser pulse energies. Because of the different reconstruction algorithms with different fields of view used for PA and US images, the PA images were cropped to enable co-registration with the US images. In particular, the skin visible in both PA and US images was used to manually co-register the images. To optimize the SNR, the co-registered sequences were divided into four sections of approximately 8 seconds that were averaged, and the averaged section with the highest acutance defined by the mean of the image gradients was used.

\subsubsection{Data annotation}
\label{sec:meth:data_anno}
After image reconstruction and post-processing, {similar to ~\cite{chlis2020sparse} and following the recommendations described in \cite{Mongan2020},} the images were manually annotated by one of three available domain experts using a standardized annotation protocol {which can be found in} Supplemental Material 1. {Eight} equally important annotation classes were distinguished during the annotation process: {blood}, skin, fat, US gel, transducer membrane, coupling agent in the transducer head (mostly heavy water), other tissue, and coupling artefact. The other tissue class was assigned to the tissue below the fat layer that does not fall into the blood category and comprises e.g. muscle tissue or conjunctive tissue. The coupling artefact class was introduced to account for a loss of signal at the edges of the image due to a lack of coupling of the transducer to the skin. Note that the distribution of annotation classes was unbalanced, meaning that the amount of pixels assigned to each class is different.

{\subsubsection{Human annotation reliability}
\label{sec:meth:IRV}
To approximate the effect of human annotation performance on our results, we performed a human annotation reliability study for the class blood, which is of particularly high clinical relevance~\cite{attia2019review}. To this end, a subset of ten test images (cf. Section \ref{sec:meth:data_split}) of one volunteer was chosen in a manner to include at least one image of every body region and body side, but otherwise random. The ten images were annotated by five domain experts and the performance of the five new annotators was assessed while using the original annotations as reference for two different metrics (cf. Section \ref{sec:validation}). To account for both the hierarchical structure and the small amount of data, a linear mixed model~\cite{ross2021can} was applied on the per image and per annotator metric values. Here we considered the body region as a fixed effect and annotator and image \textit{id} as random intercepts.}

\subsubsection{Data split}
\label{sec:meth:data_split}
The data of the 16 volunteers was divided into training/validation and test sets while respecting the underlying hierarchical structure: A training and validation set was assembled from the images of {randomly selected} ten volunteers {and split for five-fold cross-validation, where each fold used the data of two randomly chosen volunteers as the validation data (N=36) and the remaining data as the training data (N=144)}. The test set comprised 108 images from the remaining {six} volunteers that were neither included in the training nor the validation set. {This data split was chosen to ensure a large number of training data while retaining a small validation data set for hyper-parameter tuning as well as to enable statistically more reliable conclusions on the test set with N>5.
}

\subsection{Segmentation methods}
\label{sec:meth:dl}
\begin{figure}[h!tb]
    \centering
    {
    \includegraphics[width=\columnwidth]{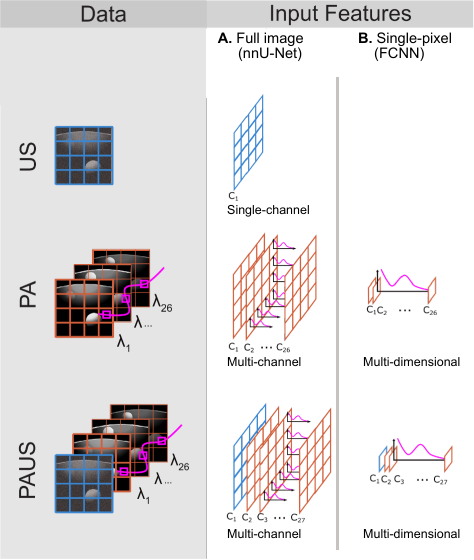}}
    \caption{\textbf{Overview of the data/network configuration.} \textit{(Left)} \textit{(top)} ultrasound (US) data (blue), \textit{(center)} multispectral photoacoustic (PA) data (orange), and \textit{(bottom)} a combination thereof (PAUS) were used as \textit{(right)} input sources for two neural network architectures: \textit{(A)} the {nnU-Net} and \textit{(B)} the fully-connected neural network (FCNN). Multispectral information encoded in PA data is indicated in pink. The input of the {nnU-Net} is a full image with one single channel for US data and multi-channel input features for PA (26 channels) and PAUS (27 channels) data. In contrast, the FCNN leverages single-pixel multi-dimensional input features for PA (26 dimensions) and PAUS (27 dimensions) data.  }
    \label{fig:overview_detailed}
\end{figure}

Due to its breakthrough successes in various fields of research and practice, we based our segmentation method on deep learning. In this context, we investigated two well-known neural network architectures with complementary strengths: the U-Net and the FCNN. 
The main difference between these architectures is the different representation of the input data they are based on. As the input representation of the {nnU-Net} is the full image where the wavelength dependency is assigned to different input channels, the network leverages local context by inherent convolutional kernels. In contrast, the input data of the FCNN is represented by single-pixel spectra. This allows for a pixel-wise classification that is completely independent of the spatial distribution of the different classes (cf. Figure \ref{fig:overview_detailed}). \\
Details of the U-Net (cf. Section~\ref{sec:UNet}) and the FCNN (cf. Section~\ref{sec:FCNN}) are presented in the following sections.

\subsubsection{Details of the {nnU-Net}} 
\label{sec:UNet}
In a recent literature review~\cite{grohl2021deep}, we found the U-Net to be the most commonly used and successful network architecture for deep learning applications in PAI{~\cite{ grohl2021learned,zhang2021deep}}. Compared to previous architectures used for image segmentation, it requires fewer training images, yields less blurred segmentation results, and is thus particularly well-suited for medical applications {\cite{ronneberger2015u}}. \\ 
{We applied the nnU-Net framework\footnote{\url{https://github.com/MIC-DKFZ/nnUNet}}, the currently best-performing framework across numerous biomedical segmentation challenges \cite{isensee2021nnu}. The core idea of the nnU-Net framework is that not the network architecture but the detailed design choices (e.g. batch size, patch size, augmentation, ensembling of folds, etc.) are key to performance optimization. In this study, we used the 2D nnU-Net configuration, since initial results showed that this network architecture performs best on our validation data. The network architecture is based on the original U-Net design~\cite{ronneberger2015u} with minor changes, such as strided convolutional downsampling layers instead of maxpooling layers. All details are described in \cite{isensee2021nnu} and a schematic of the network architecture can be found in Figure A.1 of the Supplemental Material 2.} \\
The size of the input layer was chosen according to the full image size (256 $\times$ 128) with channels $N_{in}$ that were assigned to the acquired wavelengths (i.e., 1 for US, 26 for PA, and 27 for PAUS), as also defined in the FCNN section. The size of the output layer was defined as the full image size (256 $\times$ 128) with 9 output channels $N_{out}$ that corresponded to the one-hot encoded representation of the eight annotation classes and one background class. {The {nnU-Net} was trained in a five-fold cross-validation and the estimations were ensembled. The loss was defined as the sum} of the Cross-Entropy {(CE)} Loss and the Soft Dice Loss. {The CE Loss of the one-hot encoded estimated data X and reference data Y is defined as:}

\begin{equation}
\begin{aligned}
{L_{CE}(X,Y) = mean(\{l_1,\dots,l_N\}^\top),}\\
{\textit{with \hspace{1mm}} l_n = - w_{y_n} \log \frac{\exp(x_{n,y_n})}{\sum_{c=1}^C \exp(x_{n,c})},}
\end{aligned}
\end{equation}
{where $w_{y_n}$ is a class-specific weighting factor, here set to 1, $N$ is the minibatch size, and $C$ the number of classes.
The Soft Dice Loss, here calculated per minibatch, is defined as:
\begin{equation}
    L_{SoftDice}(X,Y)= \frac {2|X\cap Y| + \epsilon_{smooth}}{|X|+|Y|+\epsilon_{smooth} +1e^{-8}}
\end{equation} 
where X and Y are the estimated and reference data, respectively, and $\epsilon_{smooth}$ a smoothing factor, here set to $1e^{-5}$.}

\subsubsection{Details of the FCNN} 
\label{sec:FCNN}
Another popular network design is the FCNN. In the context of PA{I}, FCNNs are well-suited for working with single-pixel spectra as input, which increases the number of training samples and makes it easy to work with sparsely annotated data.\\

The FCNN {architecture is based on a previous publication~\cite{grohl2021learned} and} consisted of an input layer of size 1 $\times$ 1 with dimensions $N_{in}$ that, as for the {nnU-Net}, corresponded to the measured wavelengths (i.e., 26 for PA, and 27 for PAUS). The input layer was followed by {an upscaling layer and a Tanh activation function, }four hidden layers of channel size ${4} \times N_{in}$ and a one-hot encoded output layer of dimension size $N_{out} = 8$. Dropout layers (20\%) and leaky ReLUs were used in between {the four hidden layers}. A diagram of the network architecture can be found in {Figure A.2 in the Supplemental Material 2.} For training of the FCNN, we used a Soft Margin Loss~\cite{pytorch} {defined as:
\begin{equation}
\begin{aligned}
    L_{SoftMargin}(X,Y)= mean(\{l_1,\dots,l_N\}^\top),\\
    \textit{with \vspace{1mm}} l_n = \sum_{c=1}^C \frac{\log(1+\exp{(-y_{c} \cdot x_{c})})}{C}
\end{aligned}
\end{equation}
where X and Y define the one-hot encoded estimated and reference data (containing -1 or 1), N the number of pixels per batch, and C is the number of classes.}
Same as the {nnU-Net}, the FCNN was implemented in Pytorch~\cite{pytorch}{, trained in a five-fold cross-validation and the estimations were ensembled. The} hyper-parameter optimization was based on a grid-search. We used a learning rate of {$1e^{-5}$}, a batch size of $1e^{4}$, 1000 batches per epoch and trained the network for 200 epochs.

\newpage
\subsection{{Performance assessment}}
\label{sec:validation}

We systematically compared the performances of different algorithm/data combinations using distance-based and overlap-based segmentation performance metrics. Specifically, we used the following metrics as {recommended in~\cite{reinke2021common} and applied in~\cite{antonelli2021medical}}:\\

 \noindent\textbf{Dice Similarity Coefficient (DSC)}\footnote{\label{monai_ft}\url{https://docs.monai.io/en/latest/metrics.html}.}~\cite{dice1945measures}
     {which is defined as: 
     \begin{equation}
        {\displaystyle DSC={\frac {2|X\cap Y|}{|X|+|Y|}}}
    \end{equation}}
     where Y is the reference and X the estimation of a label class.\\ 
     
 \noindent\textbf{ Normalized Surface Distance  (NSD)}~\cite{nikolov} {defined as:
         \begin{equation}
        {\displaystyle NSD={\frac {|S_x\cap B_y^{\tau}| + |S_y\cap B_x^{\tau}|  }{|S_x|+|S_y|}}}
    \end{equation}
    with the tolerance $\tau$ and the surfaces $S_x$ and $S_y$ and the border regions $B_x$ and $B_y$ of the reference X and the estimation Y respectively. Here, the tolerance was set to $\tau$ = 1 pixel for all except the blood classes.} This value was chosen as the most critical value since no inter-rater {reliability} was available. {For blood, the tolerance was calculated based on the human annotation reliability analysis (cf. Section \ref{sec:meth:IRV}) as proposed in ~\cite{nikolov}. For every test image, the average nearest neighbor distances between the surface of the reference and the re-annotated blood vessel segmentations (surface distance) were calculated. The intercept of the fitted linear mixed model (cf. Section \ref{sec:meth:IRV}) was used as the NSD tolerance value $\tau_{blood}$ = 5 pixels (cf. Table A.2 in the Supplemental Material 2).}
\\

For each test image, the performance metrics were calculated per annotation class{, hence the validation results are not biased by imbalances in the amount of pixels of the different classes}. Note that the metrics, depending on the implementation, handle edge cases differently. {Here, we decided to not compute the DSC, if a class is not existent in the reference and to not compute the NSD, if a class is not existent in the reference or in the estimation.} In addition to the annotation class-specific metric values, the class-specific values were averaged per test image, which we refer to as values of \textit{all structures}. Furthermore, descriptive statistics over the test instances were computed that resulted in one average value per algorithm/data configuration, annotation class, and metric used. Note that this approach accounts for the hierarchical nature of the data.   \\

\newpage
{Additionally, for the class \textit{blood}, the overall human performance was calculated using the DSCs of the reference and the re-annotated blood vessel segmentations. In analogy to the calculation of $\tau_{blood}$, the mean and the standard deviation of the performance of human annotators were determined by applying the linear mixed model on the per image and per annotator DSC values (cf. Section \ref{sec:meth:IRV}).}\\

A comparison between the algorithms was performed using the challengeR toolkit\footnote{\url{https://phabricator.mitk.org/source/challenger/}} which is especially suited to analyze and visualize benchmarking results~\cite{wiesenfarth2021methods}. A statistical rank-then-aggregate-based approach was chosen to rank the methods, similar to the ranking scheme of the Medical Segmentation Decathlon 
~\cite{antonelli2021medical}. First, a ranking per test case and algorithm was calculated. Second, the mean of the rankings per algorithm was calculated, resulting in the final rank. We chose the DSC as the primary metric for the ranking and defined the segmentation for blood vessels, skin, and all structures as three separate tasks. The segmentation of skin and blood vessels can be considered two of the most crucial tasks for segmentation algorithms in the field of PAI~\cite{grohl2021deep}. The corresponding DSC values were aggregated across the three poses and two body sites (cf. Figure~\ref{fig:hierarchy}), resulting in N = 18 test cases per annotation class. Note that initial results of the challengeR toolkit did not show any major differences when using the NSD values or different aggregation schemes or a different (non test-based) ranking method.

% \newpage
\section{Experiments and Results}

We designed one experiment to qualitatively analyze the data annotations and two experiments to verify the following core hypotheses: 

\begin{enumerate}
    \item Automatic multi-label segmentation of PA and PAUS data is feasible (cf. Section \ref{sec:exp:feasible}).
    \item Automatic multi-label segmentation is feasible even when being applied to morphologically different test data (cf. Section \ref{sec:exp:robustness}).
\end{enumerate}

The following sections detail the experimental design and the results for each.

\begin{figure}[h!tb]
\centering
\includegraphics[width=\columnwidth]{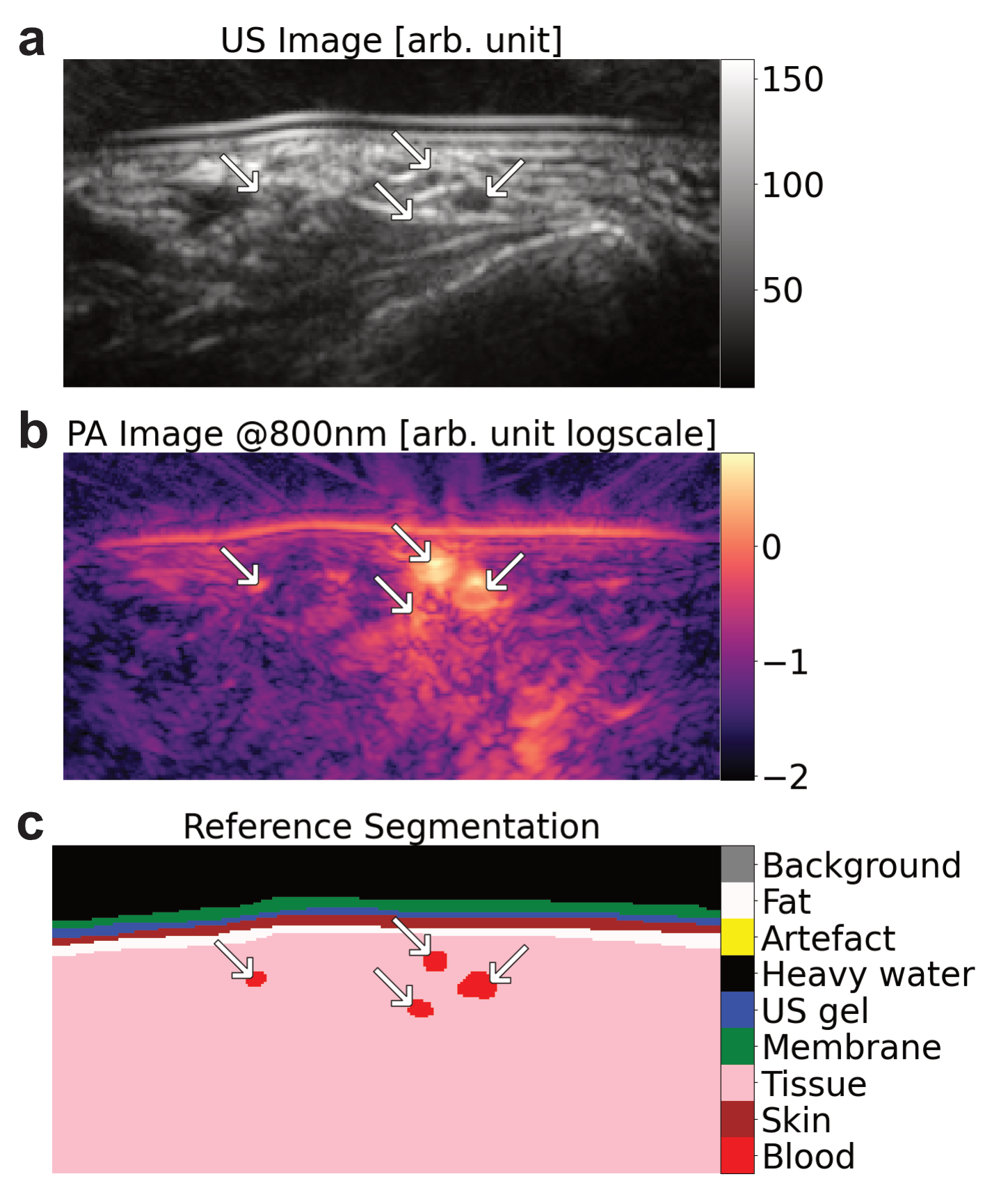}
\caption{\textbf{Example of the forearm data set.} \textit{(a)} An ultrasound (US) image, \textit{(b)} a photoacoustic (PA) image at 800\,nm, and \textit{(c)} the reference segmentation mask. The white arrows denote the location of the same vessel structures in all three images.}
\label{fig:dataexample}
\end{figure}

\vspace{1cm}

\subsection{Automatic multi-label segmentation of PA and PAUS data is feasible}
\label{sec:exp:feasible}

\begin{figure}[h!tb]
\centering
    {
\includegraphics[width=\columnwidth]{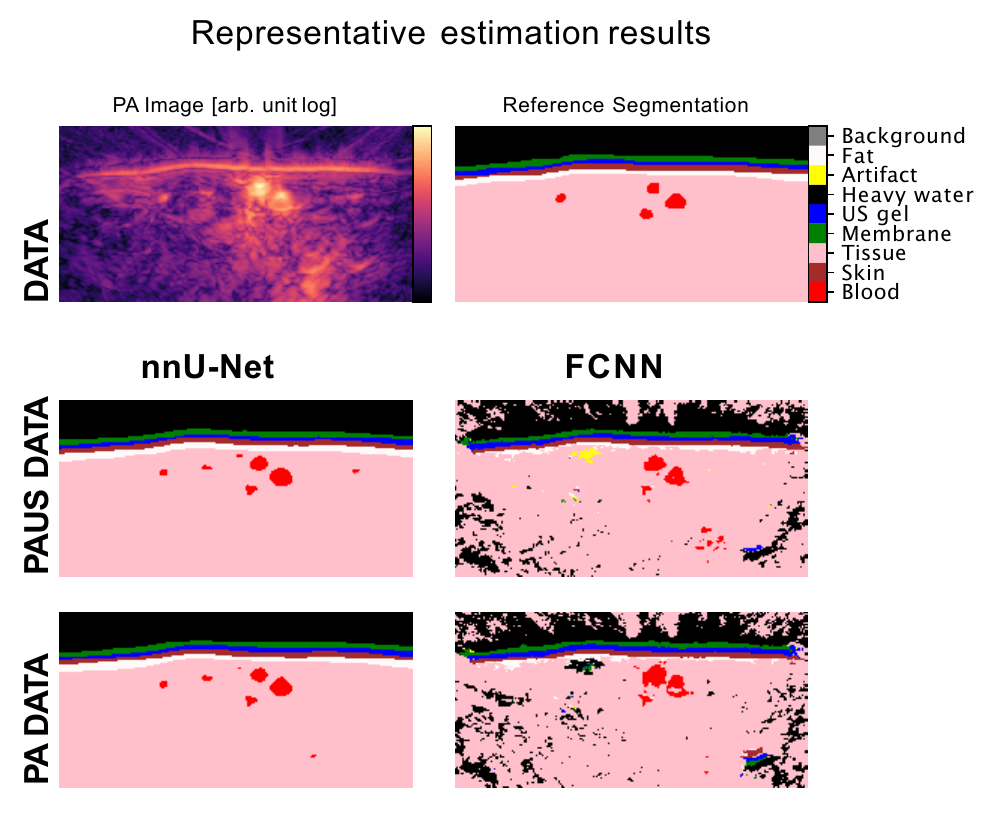}}
\caption{\textbf{Segmentation results show agreement with the reference segmentation.} A representative example image was chosen according to the median blood Dice Similarity Coefficient (DSC) (calculated on images with at least 60 pixels classified as blood) for the {nnU-Net} using photoacoustic and ultrasound (PAUS) data as input. The \textit{first row} shows the input data: \textit{(left)} the logscaled photoacoustic (PA) image at 800 nm and \textit{(right)} the reference segmentation. The estimated segmentation maps are shown below for \textit{(left)} the nnU-Net and \textit{(right)} the fully-connected neural network (FCNN) based on \textit{(second row)} PAUS input data and \textit{(third row)} PA input data.}
\label{fig:res:feasability:overview}
\end{figure}

We performed a \textit{feasibility} experiment in which we used different combinations of data and inference models to evaluate differences in performance when segmenting the labelled annotation classes. We trained the {nnU-Net} on US data only, PA data only, and a combination of both PA and US data (PAUS) {(cf. Figure \ref{fig:dataexample})}. The FCNN training was performed only on PA and PAUS data because it estimates annotation classes from pixel-wise spectra alone, and the one-dimensional nature of the US data does not provide this information. Figure~\ref{fig:res:feasability:overview} shows the estimated segmentations (having at least 60 blood pixels) corresponding to the median DSC for the blood class. While the FCNN estimations for the PA data sets qualitatively look plausible, there is much more noise in the estimated labels compared to the nnU-Net estimates.\\

\begin{figure}[h!tb]
    \centering
    {
    \includegraphics[width=\columnwidth]{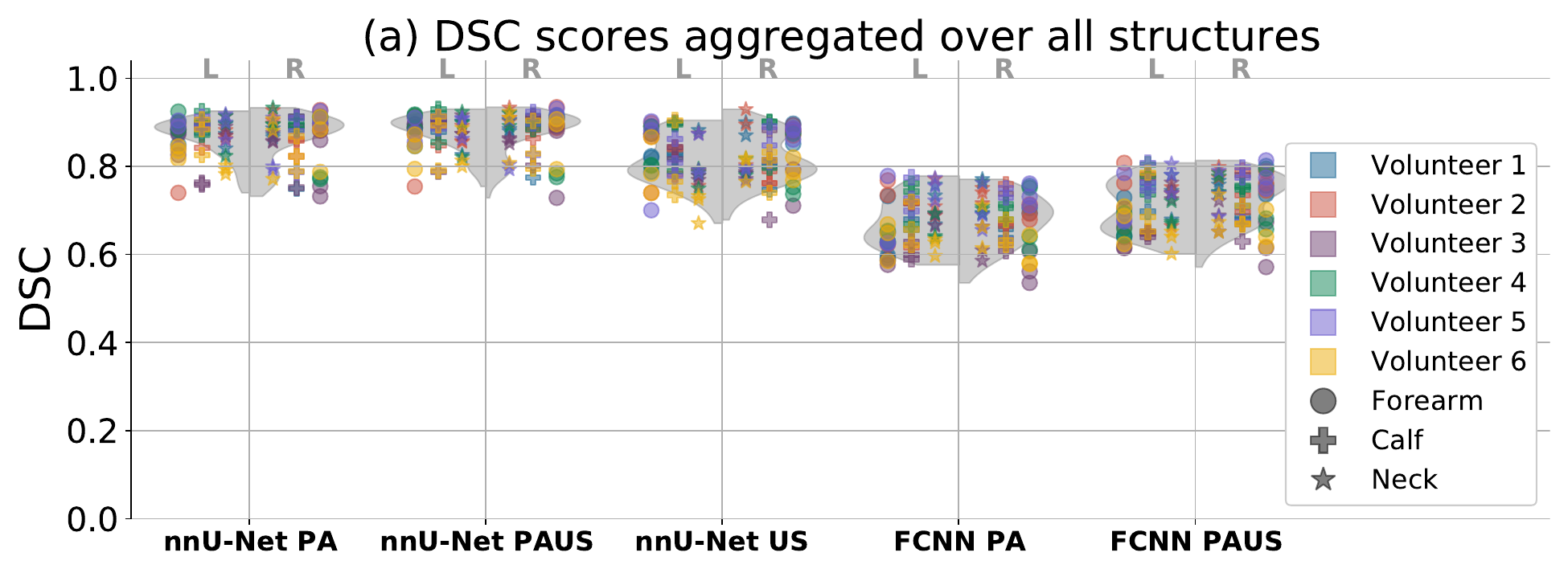}}
    \vspace*{0.25em}\\
        {\includegraphics[width=\columnwidth]{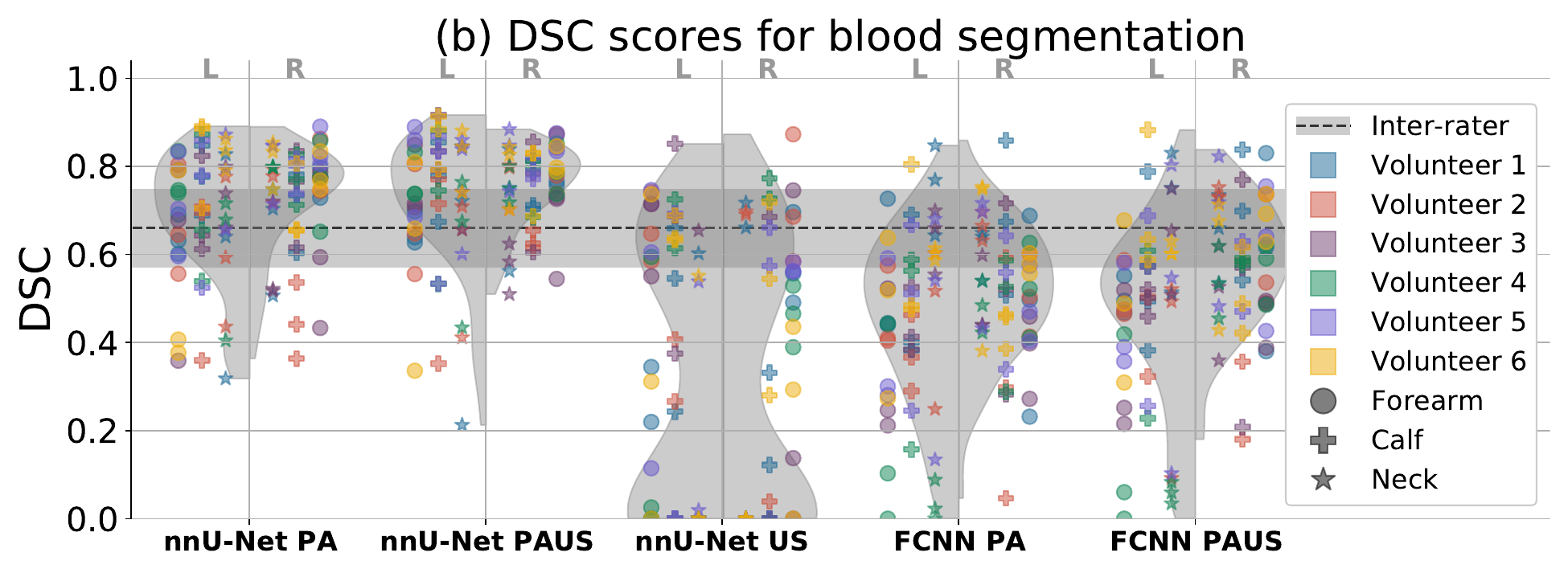}}
    \vspace*{0.25em}\\
        {\includegraphics[width=\columnwidth]{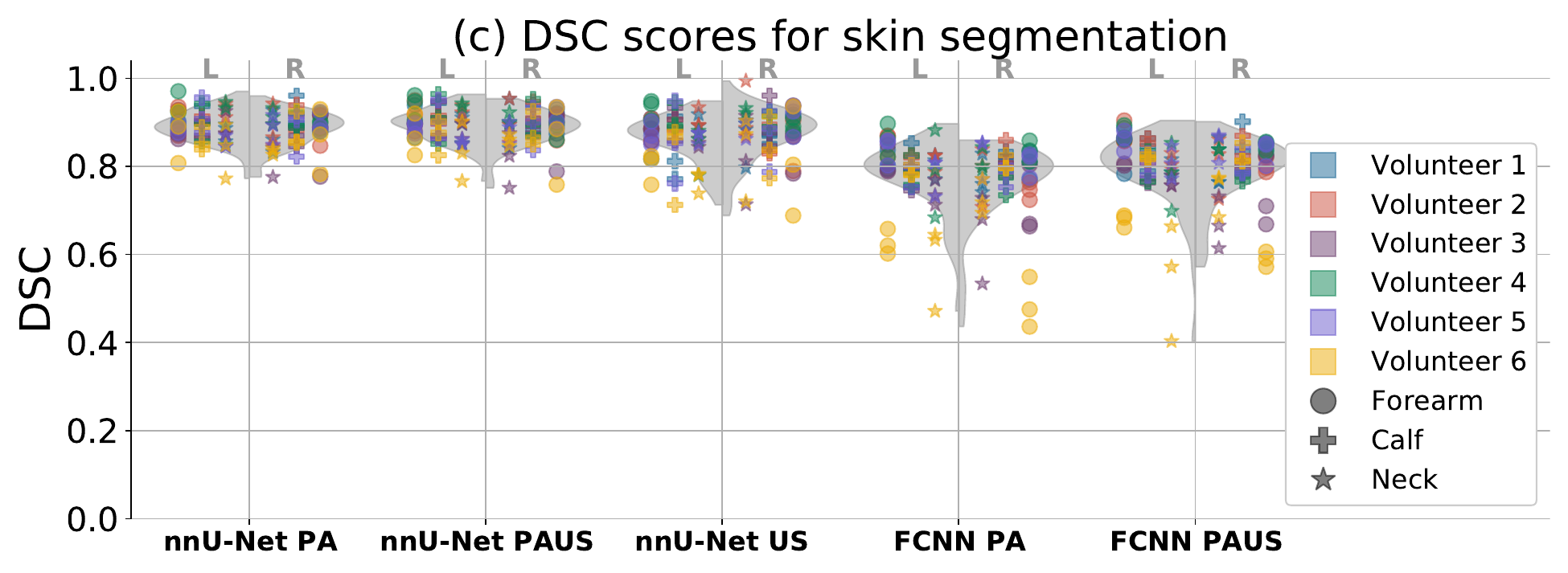}}
    \caption{\textbf{Raw data plot for all algorithm/data combinations.} Dice Similarity Coefficient (DSC) scores achieved in the feasibility experiment \textit{(a)} averaged over all structures, \textit{(b)} for blood, and \textit{(c)} for skin. A separate plot is shown for each of the algorithm/data combinations. Color/Shape coding enables distinguishing the six volunteers (six colors) as well as the different target structures (circle, plus, and star for forearm, calf, and neck respectively). Measurements from the left and right side of the volunteers' bodies are plotted at the left and right of the vertical lines, respectively. The grey density plots show the relative score frequencies separately for each side. {For blood, the mean and standard deviation of the performance of the human annotators are shown as the dotted line and the shaded area, respectively.} }
    \label{fig:feasibility_raw_data}
\end{figure}

\begin{figure}[h!tb]
    \centering
    {
    \includegraphics[width=\columnwidth]{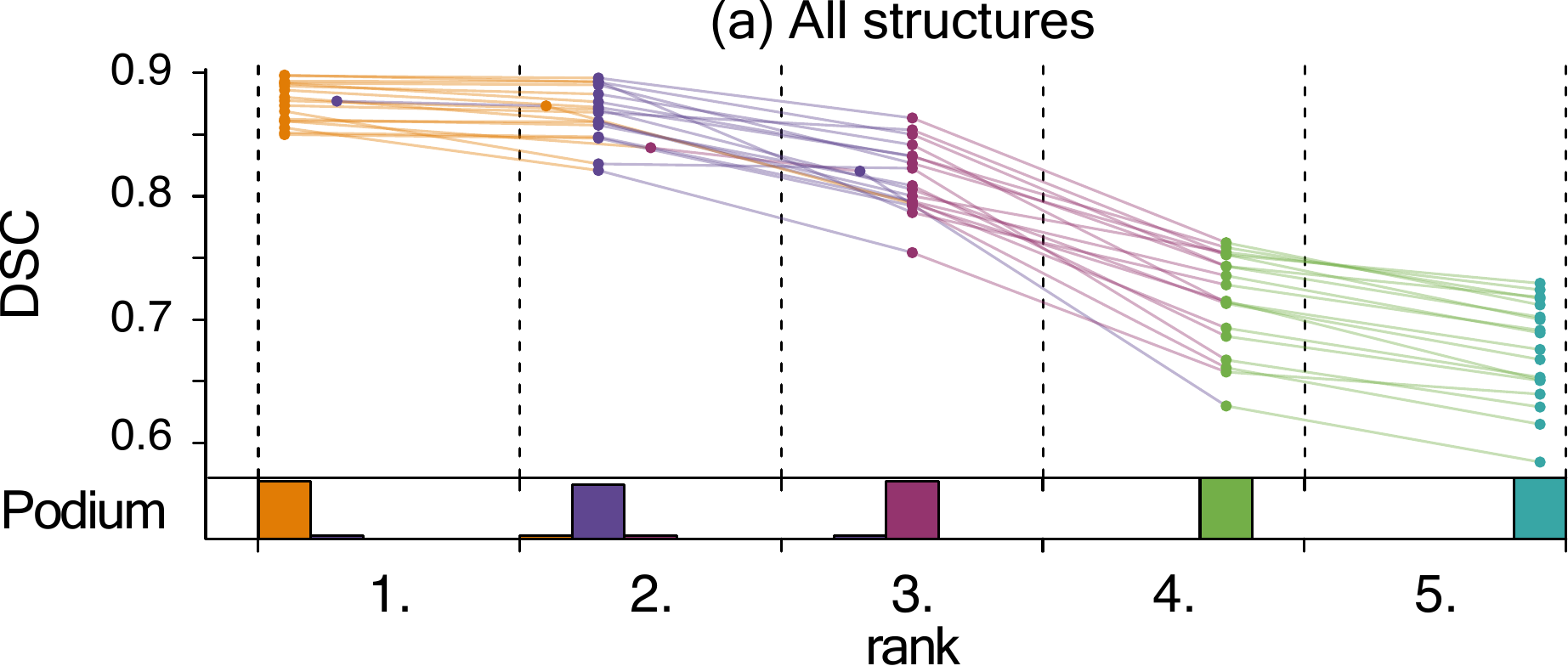}}
    \vspace*{0.25em}\\
    {
    \includegraphics[width=\columnwidth]{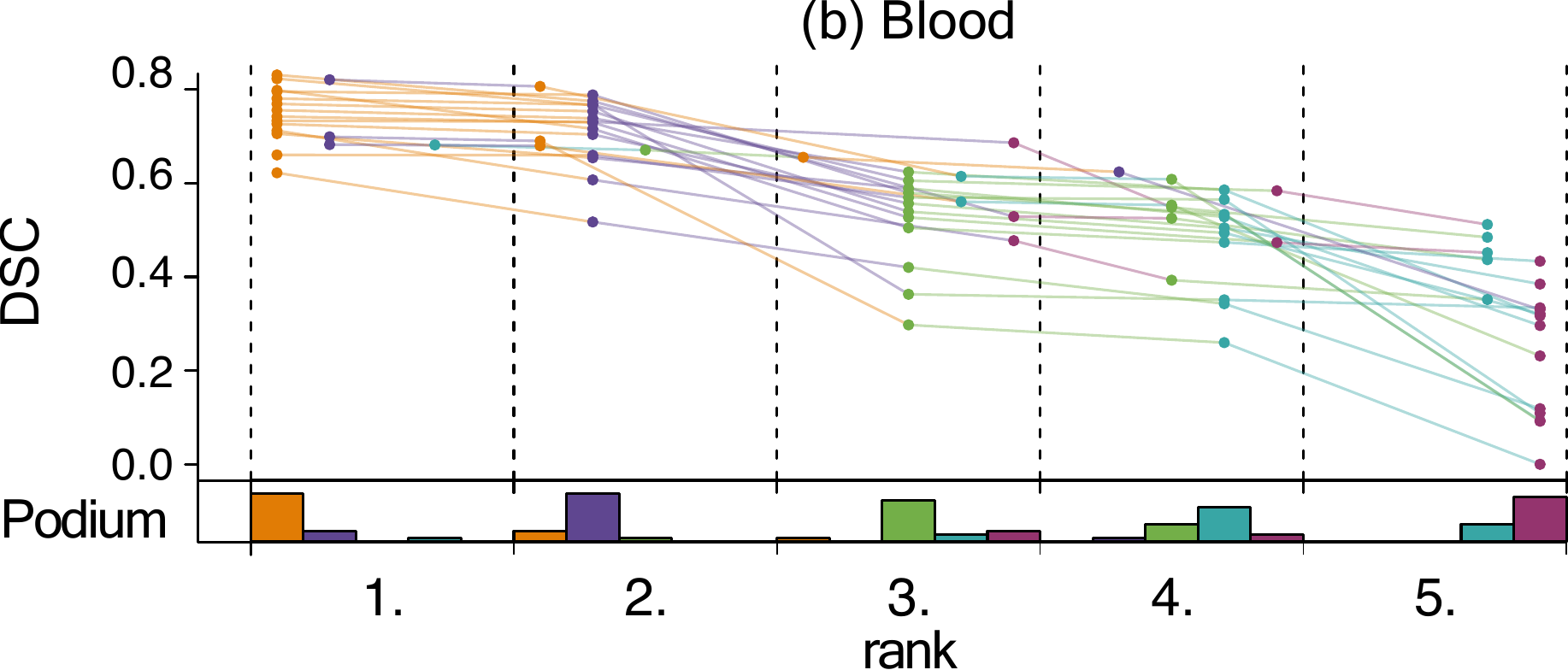}}
    \vspace*{0.25em}\\
    {\includegraphics[width=\columnwidth]{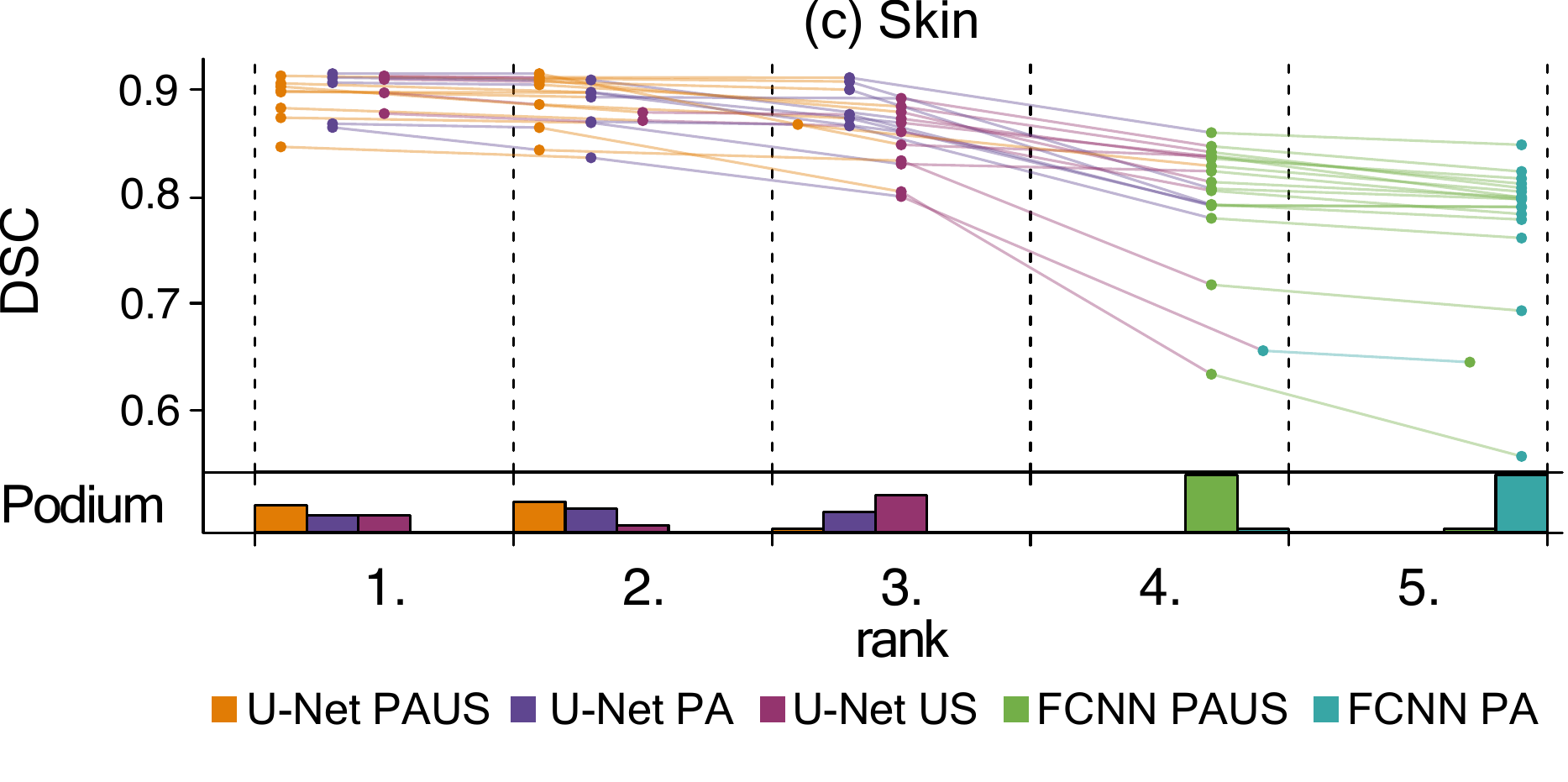}}
    \caption{\textbf{The {nnU-Net} generally outperforms the fully-connected neural network (FCNN) and when using US data only, the {nnU-Net} struggles to segment blood vessels.} For each of the raw data plots in Figure \ref{fig:feasibility_raw_data} this figure shows an accompanying podium plot generated with the \emph{challengeR} toolkit \cite{wiesenfarth2021methods} that displays the relative performance of the Dice Similarity Coefficient (DSC) aggregated across the three poses and two body sides (cf. Figure~\ref{fig:hierarchy}). \textit{(Upper parts)} Participating models and corresponding DSC values are color-coded and ordered according to the achieved ranks from best (1) to worst (5). DSC values corresponding to identical test cases are connected by a line (spaghetti structure). \textit{(Lower parts)} The bar charts represent the relative frequency at which each model achieved the respective rank.}
    \label{fig:feasibility_raw_data_podium}
\end{figure}

Table~\ref{tab:results:unet_fcnn} shows the DSC and the NSD results of all input data for the {nnU-Net} and FCNN segmentation architectures. Figure~\ref{fig:feasibility_raw_data} presents the distribution of the DSC for blood, skin, and the average over all structures. {Especially the {nnU-Net}s trained on multispectral data achieve slightly higher DSCs for blood compared to the performance of the human annotators (mean of 0.66, standard deviation of 0.09).} The results for all annotation classes can be seen { in Table A.1 and details of the fitted linear mixed model can be found in Table A.2 in the Supplemental Material 2}. The model performances differed the most in case of blood segmentation. Especially for this annotation class, the {nnU-Net} performed substantially better than the FCNN. Figure~\ref{fig:feasibility_raw_data_podium} confirms this finding and also indicates that it is beneficial to use PA or PAUS data compared to using US data as sole input for the blood segmentation. Additional qualitative results can be found {in Figure A.3} in the Supplemental Material 2. \\

\begin{table}[h!tb]
    \centering
    {
    \begin{tabular}{ p{0.4cm}p{1cm}p{0.75cm}p{0.75cm}p{0.75cm}p{0.75cm}p{0.75cm}}
    \hline
     & \textbf{Structure Type} & \textbf{{nnU-Net}} & \textbf{{nnU-Net}} & \textbf{{nnU-Net} } & \textbf{FCNN } & \textbf{FCNN } \\
    & & \textbf{PA} & \textbf{PAUS} & \textbf{ US}& \textbf{PA} & \textbf{PAUS} \\
    \hline
    % $\leftarrow$\\
    {} &\\
    \cellcolor{white}{} &   All &  0.83 &  0.85 &  0.80 &  0.62 &  0.66 \\
    \rowcolor{bloodred!50}
    \cellcolor{white}{} & Blood &  0.71 &  0.74 &  0.32 &  0.48 &  0.53 \\
    \rowcolor{skinbrown}
    \cellcolor{white}{} &  Skin &  0.89 &  0.89 &  0.87 &  0.77 &  0.79 \\
    {\multirow{-6}{*}{\rotatebox[origin=c]{90}{\parbox[c]{1cm}{\centering \textbf{DSC}}}}} & \\
    \hline
    % $\leftarrow$\\
    {} &\\
    \cellcolor{white}{} &  All &  0.88 &  0.89 &  0.84 &  0.59 &  0.61 \\
    \rowcolor{bloodred!50}
    \cellcolor{white}{} & Blood & 0.84 &  0.85 &  0.47 &  0.75 &  0.75 \\
    \rowcolor{skinbrown}
    \cellcolor{white}{} & Skin &  0.98 &  0.98 &  0.97 &  0.87 &  0.89 \\
    {\multirow{-5}{*}{\rotatebox[origin=c]{90}{\parbox[c]{2cm}{\centering \textbf{NSD}}}}} & \\
    \hline
    \\
    \end{tabular}}
    \caption{\textbf{The mean performance scores show the feasibility of the method across multiple performance metrics.} The Dice Similarity Coefficients (DSCs) and Normalized Surface Distances (NSDs) for the estimation results achieved by the {nnU-Net} and the fully-connected neural network (FCNN) leveraging photoacoustic (PA) data, ultrasound (US) data, and a combination thereof (PAUS), were calculated over all structures and for blood and skin separately. Higher DSC and NSD values are better.}
    \label{tab:results:unet_fcnn}
\end{table}

\newpage
\subsection{Automatic multi-label segmentation is feasible even when applied to morphologically different test data}
\label{sec:exp:robustness}

To investigate the \textit{robustness} of the proposed approach with respect to morphologically different target structures, we conducted experiments in which we used different target structures for training and testing of our algorithm. Specifically, we trained the networks exclusively on (A) the neck and calf, (B) the forearm and neck, and (C) the forearm and calf measurements of the training set and estimated the annotation classes on (A) the forearm, (B) the calf, and (C) the neck measurements of the test set. The different combinations of data and inference models were analogous to the feasibility experiments.\\

\begin{figure}[htb]
    \centering
    {
    \includegraphics[width=1.1\columnwidth]{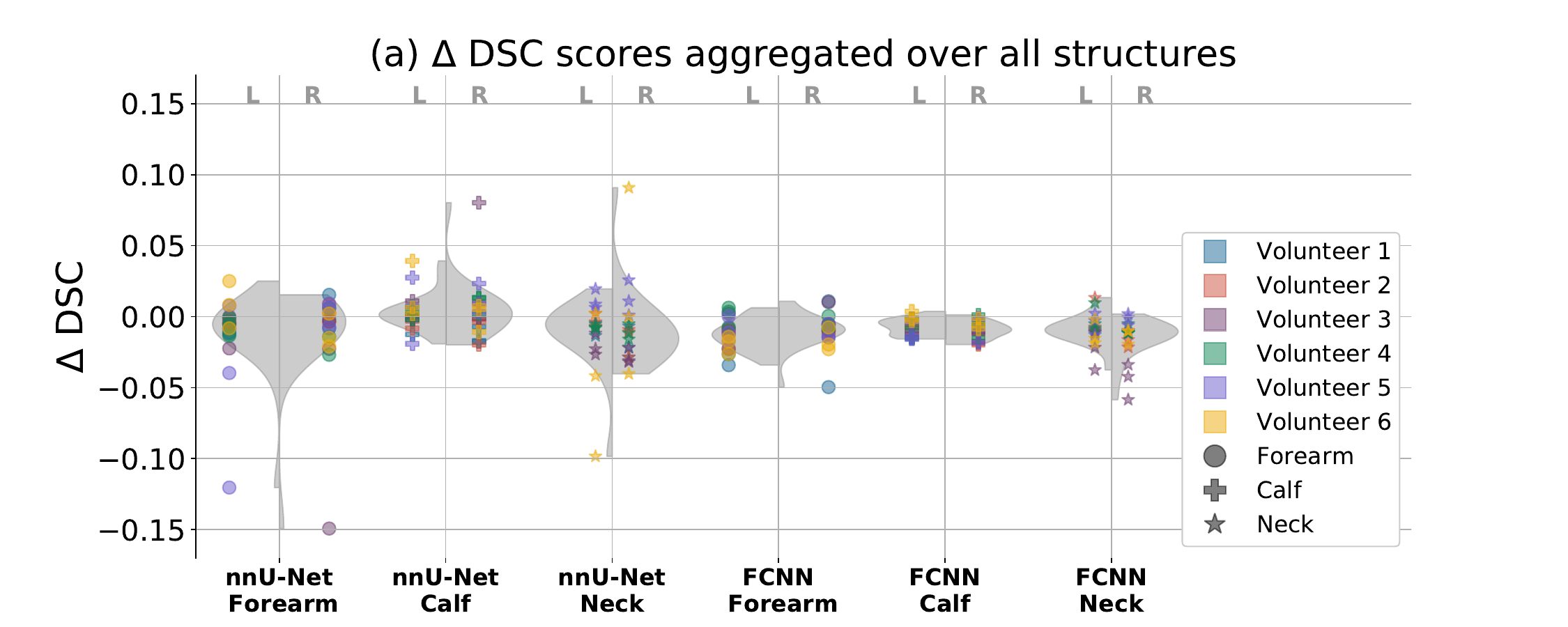}}
    {\includegraphics[width=1.1\columnwidth]{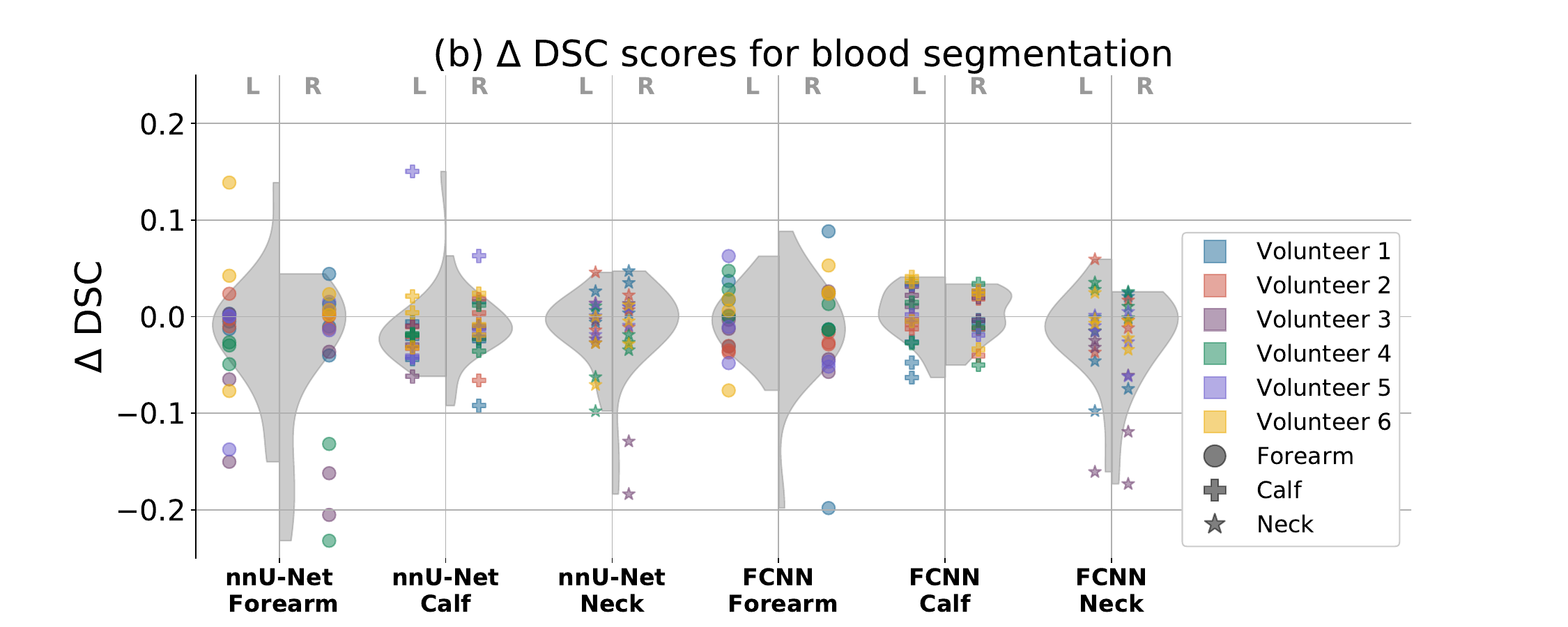}}
    {\includegraphics[width=1.1\columnwidth]{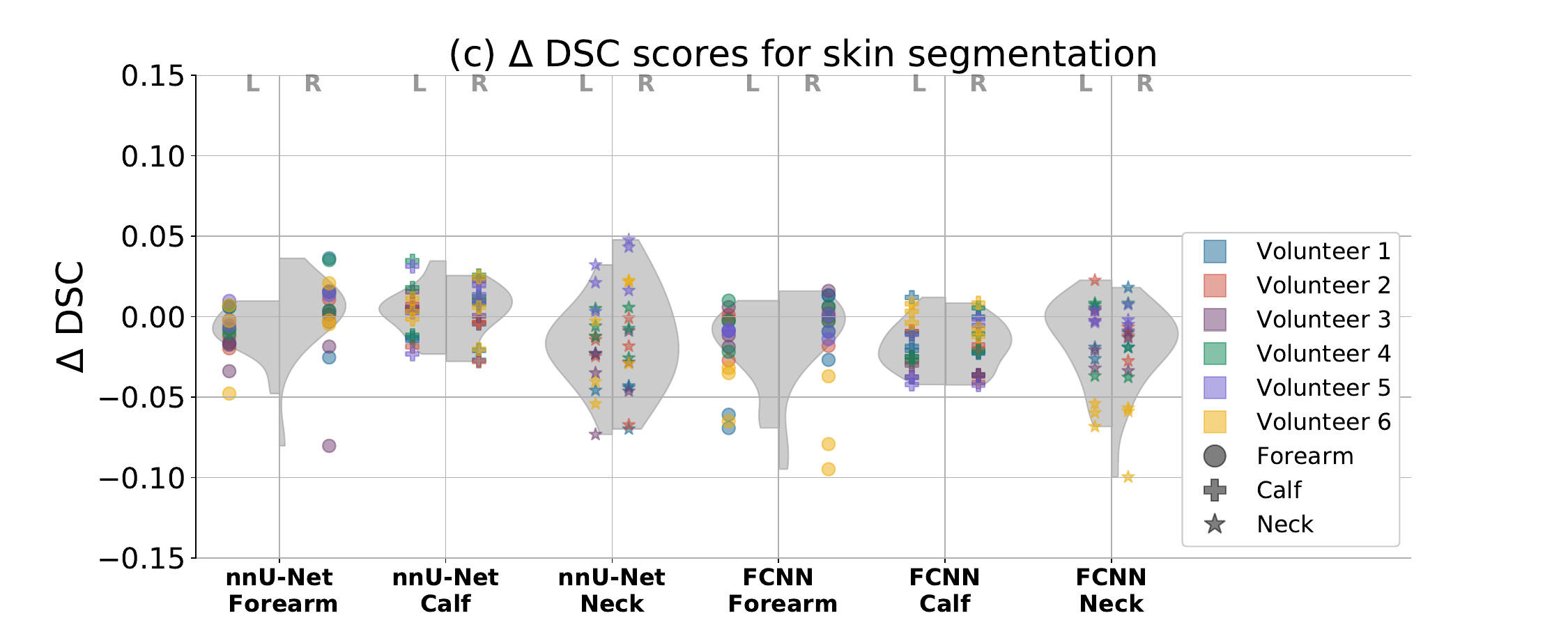}}
    \caption{\textbf{The fully-connected neural network (FCNN) tends to be more robust to body sites that were not included in the training data compared to the {nnU-Net}.} The respective differences of the Dice Similarity Coefficient (DSC) results \textit{(a)} averaged over all structures, \textit{(b)} for skin , and \textit{(c)} for blood between the robustness experiments and the feasibility experiment for the {nnU-Net} and FCNN trained on photoacoustic and ultrasound (PAUS) data are shown. Values higher than zero denote an improved performance of the robustness results compared to the feasibility results. Values smaller than zero indicate the opposite. The separate plots for each of the training/test data combinations are indicated by the respective test set: \textit{forearm} models trained on calf and neck data and tested on forearm measurements, \textit{calf} models trained on forearm and neck data and tested on calf measurements, and \textit{neck} models trained on forearm and calf data and tested on neck measurements. Color/Shape coding enables distinguishing the six volunteers (six colors) as well as the different target structures (circle, plus, and star for forearm, calf, and neck respectively). Measurements from the left and right side of the volunteers' bodies are plotted to the left and right of the vertical lines, respectively. The grey density plots show the relative score frequencies separately for each side. The frequency distributions of the FCNN differences are narrower compared to those of the {nnU-Net}.}
    \label{fig:robustness_raw_data_vs_feas}
\end{figure}

The DSC and the NSD scores for the baseline method as well as the scenarios (A - C) are shown in Table~\ref{tab:res:robustness}. In general, the {nnU-Net} models outperform the FCNN models. However, the differences of DSC results between the robustness and feasibility experiments, as shown in Figure~\ref{fig:robustness_raw_data_vs_feas}, indicate that the FCNN is more robust to body sites that are not included in the training data set compared to the {nnU-Net}.
Additional qualitative as well as quantitative results can be found {in Figures A.4 - A.6 and Tables A.3 - A.5} in the Supplemental Material 2. 

\begin{table*}[htb]
\centering
{
  \begin{tabular}{ p{2cm}p{2cm}p{2.3cm}p{2.3cm}p{2.3cm}p{2.3cm}}
  \multicolumn{5}{l}{\textbf{Robustness results on PAUS data}.}\\  \hline
  \\
  Scenario & & & \centerline{\textbf{A}}&\centerline{\textbf{B}}& \centerline{\textbf{C}}\\
  \\
   Training data  &   & \textbf{All} (Baseline) & \textbf{Neck/Calf} & \textbf{Forearm/Neck} & \textbf{Forearm/Calf} \ \\ \\            
   Test data      &   & \textbf{All} (Baseline) & \textbf{Forearm} & \textbf{Calf} & \textbf{Neck} \\                 
  \\ 
  \\
  & \textbf{Structure Type} & \textbf{{nnU-Net}/FCNN} & \textbf{{nnU-Net}/FCNN} &\textbf{{nnU-Net}/FCNN} & \textbf{{nnU-Net}/FCNN}
  \\
\hline
%   $\leftarrow$\\
{{}} &\\
\cellcolor{white}{} &  All &  0.85 / 0.66 &  0.82 / 0.65 &  0.86 / 0.66 &  0.83 / 0.64\\
\rowcolor{bloodred!50}
\cellcolor{white}{} &  Blood &  0.74 / 0.53 & 0.70 / 0.52 &  0.74 / 0.54 &  0.72 / 0.49\\
\rowcolor{skinbrown}
\cellcolor{white}{} &  Skin &  0.89 / 0.79 &  0.86 / 0.75 &  0.90 / 0.80 &  0.89 / 0.78\\
{\multirow{-6}{*}{\rotatebox[origin=c]{90}{\parbox[c]{1.4cm}{\centering \textbf{DSC}}}}} & \\
\hline
%  $\leftarrow$\\
{{}} &\\
\cellcolor{white}{} &  All &  0.89 / 0.61 &  0.87 / 0.59 &  0.89 / 0.61 &  0.88 / 0.60\\
\rowcolor{bloodred!50}
\cellcolor{white}{} &  Blood &  0.85 / 0.75 &  0.84 / 0.68 &  0.83 / 0.78 & 0.85 / 0.75\\
\rowcolor{skinbrown}
\cellcolor{white}{} &  Skin &  0.98 / 0.89 &  0.97 / 0.87 &  0.98 / 0.86 &  0.98 / 0.89\\
{\multirow{-6}{*}{\rotatebox[origin=c]{90}{\parbox[c]{1.4cm}{\centering \textbf{NSD}}}}} & \\
\hline
\\
{} &\\
\end{tabular}}

\caption{\textbf{Semantic segmentation results can differ on geometrically different test images when using a combination of photoacoustic and ultrasound (PAUS) data.} The baseline feasibility experiment is compared to the robustness experiment results of semantic segmentation tested on images from geometries (forearm, calf, and neck) that were not included in the training set. We use the Dice Similarity Coefficient (DSC) as well as the Normalized Surface Distance (NSD). Higher DSC and NSD values are better.}
\label{tab:res:robustness}
\end{table*}

\clearpage
\section{Discussion}

To our knowledge, this paper is the first to show that fully-automatic \emph{multi-label} semantic image annotation of PA images with deep learning is feasible. {We designed two experiments around one core hypothesis that questioned: \textbf{1)} the general feasibility of multi-label image segmentation based on PA images and \textbf{2)} the robustness towards changes in tissue geometry. We found that the task is generally feasible across different neural network architectures {and that, compared with pure US images, the multispectral nature of PA images is particularly beneficial for blood segmentation (DSC of PAUS nnU-Net was 0.74, DSC of US {nnU-Net} was 0.32)}. To this end, we used two distinct types of networks throughout the experiments:} an FCNN that utilizes single-pixel spectral information and a {nnU-Net} that additionally incorporates spatial context information. {Specifically, we applied the nnU-Net}, the current state of the art in medical image segmentation. {On PAUS data, averaged across all annotation classes, the nnU-Net achieved a DSC of 0.85 and the FCNN of 0.66.} Furthermore, the method was robustly applicable even to data where samples of the imaged body site {were} not included in the training data.\\

In the feasibility experiment, the {nnU-Net} trained on multispectral data achieved better results on average compared to the FCNN. We attribute this to its ability to take into account the spatial image context. However, even though the FCNN was only able to estimate the labels based on single-pixel spectra, the resulting images were plausible as well. The overall results of the networks were very convincing, achieving high overlap-based and contour-based metric values for the majority of classes. The worst performances were achieved for the blood and coupling artefact classes, for which the most obvious explanation is the difficulty of annotating these areas. Nevertheless, the spectra of the annotation classes seem to be very characteristic for the respective class. {
In an additional experiment, we noticed that the nnU-Net trained on PAUS data performs only marginally worse when using five wavelengths evenly sampled from the 26 wavelengths. This leads us to assume that, provided the characteristic spectra are sufficiently sampled, fewer wavelengths can be leveraged.   
{A systematic analysis of failure cases (representative image shown in Figure A.3 in the Supplemental Material 2) revealed two common sources of error: (1) over-segmentation of small superficial vessels and (2) segmentation of regions of high intensity that were not identified as vessels by the annotators.}
} \\

In our robustness experiment, the performance of the networks that were trained on PAUS data decreased slightly in comparison to the feasibility experiment {in most of the cases}. We suspect the domain shift between different body regions that were explicitly included in this experiment to be a reason for this finding. This indicates that the variations between different body regions might have an impact on the algorithm performance although the tissue composition of the regions investigated was similar. It can be expected that the performance would degrade even further for vastly different structures (e.g. in abdominal surgery) or when being applied to a different cohort, e.g. cancer patients. It has to be noted that the test data set of the robustness experiment compared to the feasibility experiment was smaller in size by a factor of three, which may also have caused the {overall} decrease in performance. {Moreover, the smaller test set size could be a reason why the results of the robustness experiments partly outperformed those of the feasibility experiment.} In future work including more data, the significance of the performance difference should be addressed. The experimental results could be a further indicator of the high potential of FCNNs for robust interpretation{, which, in this study, is referred to as an independence of neural networks of morphologically different training and test data,} as the distribution of DSC differences between the robustness and feasibility results was narrower than that of the {nnU-Net}. Compared to convolutional approaches, FCNNs are by design more independent of morphological variations within the training and target domains. Ensembling strategies based on the estimation uncertainty using both the {nnU-Net} and FCNN results{, leveraging the FCNN results as a prior, or post-processing of the resulting images} might be able to combine the respective advantages of both.\\

In the broader context of biomedical image analysis, it was suggested that model performance can be hampered by the quality of the reference segmentation labels~\cite{zlateski2018importance}. The main problem with manual labelling is that the process is time-consuming, {requires expert knowledge}, and is error-prone. {To reduce the differences between annotations, we devised a structured annotation protocol for standardization of the labelling process and facilitated differentiation of various structures (cf. Supplemental Material 1). {A} leave-one-out cross-validation of all acquired images {did not} reveal any obvious differences between the segmentation masks of the individual volunteers drawn by different annotators.} Still, there remains ambiguity in the images, for example when delineating the apparent size {and location} of blood vessels in the PA signal{, as highlighted in an additional human annotation reliability  analysis. Especially the slightly higher blood DSC results of the nnU-Net trained on multispectral data compared to the performance of human annotators indicate that the networks might be able to replace manual labeling - if the annotation quality is high enough.} In some images, we have found size and position mismatches between the PA and US images, which might be introduced by differences in the speed of sound across different volunteers or by a blurring of the PA signals, which can be attributed to the limited bandwidth of the US detection elements, their impulse response, as well as artefacts introduced by the reconstruction algorithm. These inherent ambiguities might be resolvable by multi-modal image registration~\cite{ren2021feasibility} or by capturing 3D instead of 2D images to exploit the spatial context information. However, obtaining high-quality 3D reconstructions with handheld linear transducers is usually very time-consuming and requires the use of additional hardware~\cite{holzwarth2020tattoo}.

While our study indicates that \emph{multi-label} semantic annotation of PA and/or US images with deep learning is feasible, the results should be interpreted with care. The biggest limitation of our study is the fact that we had a very low number of test images. With only 108 images from 6 volunteers, no broad conclusions should be drawn from the results, especially with respect to the relative performance of the different architecture/input combinations. A related potential problem is the hierarchical nature of the data set (16 volunteers, three imaging sites, three images per site, left and right side of the body), which complicates a rigorous statistical analysis. In fact, based on the data visualization in Figure~\ref{fig:feasibility_raw_data}, we found no clear evidence that the algorithms performed differently on test images of the left and right side of the body. At the same time, in some cases there is a clustering of DSC results within the same site. Our test data therefore cannot be regarded as independent, which was the reason for us to report the mean performance without standard deviation in Tables~\ref{tab:results:unet_fcnn} and~\ref{tab:res:robustness}. Additionally, our experimental design may have led to a bias favoring networks trained on PAUS data. On the one hand, the manual annotations were done using both the PA and US images. On the other hand, the number of learnable parameters of the networks increased with the number of input channels (in this case corresponding to the number of input spectra), which may improve the performance of the corresponding networks \cite{tan2019efficientnet}. Moreover, the presented work is limited to the given annotated classes. {In this paper, we categorized pixels into eight classes. Future work with more training data could investigate semantic scene segmentation with hierarchical class structures, for example to differentiate arteries and veins within the class blood. It should be mentioned that initial experiments on distinguishing the two blood classes did not yield high performance with the limited training data per class that we currently have.} In future work, this method could be extended to segment specific structures that are oncologically relevant. However, these structures would need to be located sufficiently close to the skin and show a characteristic PA spectrum.\\

Overall, our work indicates that neural network-based semantic image segmentation {of multispectral PA images }is feasible, producing robust estimates even with relatively small amounts of training data. {We believe that algorithms for automatic analysis of photoacoustic images are an important step towards clinical translation as they can assist physicians in understanding multispectral photoacoustic images. Especially in combination with {wavelength-dependent} tools for functional parameter estimation, such as blood oxygen saturation, they allow for the creation of powerful and clinically impactful visualizations of the imaged tissue structures.}\\

\newpage
\section*{Additional Information}
\label{sec:add_info}
The healthy human volunteer experiments were approved by the ethics committee of the medical faculty of Heidelberg University under reference number S-451/2020 and the study is registered with the German Clinical Trials Register under reference number DRKS00023205.

\section*{Acknowledgements}
This project was funded by the European Research Council (ERC) under the European Union’s Horizon 2020 research and innovation programme through the ERC starting grant COMBIOSCOPY (grant agreement No. ERC-2015-StG-37960) and consolidator grant NEURAL SPICING (grant agreement No. [101002198]) and the Surgical Oncology Program of the National Center for Tumor Diseases (NCT) Heidelberg.\\
Part of this work was funded by Helmholtz Imaging (HI), a platform of
the Helmholtz Incubator on Information and Data Science.

\printcredits


\begin{thebibliography}{2}
\providecommand{\natexlab}[1]{#1}
\providecommand{\url}[1]{\texttt{#1}}
\expandafter\ifx\csname urlstyle\endcsname\relax
  \providecommand{\doi}[1]{doi: #1}\else
  \providecommand{\doi}{doi: \begingroup \urlstyle{rm}\Url}\fi

\bibitem[Isensee et~al.(2021)Isensee, Jaeger, Kohl, Petersen, and
  Maier-Hein]{isensee2021nnu}
F.~Isensee, P.~F. Jaeger, S.~A. Kohl, J.~Petersen, and K.~H. Maier-Hein.
\newblock nn{U-Net}: a self-configuring method for deep learning-based
  biomedical image segmentation.
\newblock \emph{Nature Methods}, 18\penalty0 (2):\penalty0 203--211, 2021.

\bibitem[Ronneberger et~al.(2015)Ronneberger, Fischer, and
  Brox]{ronneberger2015u}
O.~Ronneberger, P.~Fischer, and T.~Brox.
\newblock U-{Net}: Convolutional networks for biomedical image segmentation.
\newblock In \emph{International Conference on Medical Image Computing and
  Computer-Assisted Intervention}, pages 234--241. Springer, 2015.

\end{thebibliography}


\begin{thebibliography}{38}
\providecommand{\natexlab}[1]{#1}
\providecommand{\url}[1]{\texttt{#1}}
\expandafter\ifx\csname urlstyle\endcsname\relax
  \providecommand{\doi}[1]{doi: #1}\else
  \providecommand{\doi}{doi: \begingroup \urlstyle{rm}\Url}\fi

\bibitem[Antonelli et~al.(2021)Antonelli, Reinke, Bakas, Farahani, Landman,
  Litjens, Menze, Ronneberger, Summers, van Ginneken,
  et~al.]{antonelli2021medical}
M.~Antonelli, A.~Reinke, S.~Bakas, K.~Farahani, B.~A. Landman, G.~Litjens,
  B.~Menze, O.~Ronneberger, R.~M. Summers, B.~van Ginneken, et~al.
\newblock The medical segmentation decathlon.
\newblock \emph{arXiv preprint arXiv:2106.05735}, 2021.

\bibitem[Attia et~al.(2019)Attia, Balasundaram, Moothanchery, Dinish, Bi,
  Ntziachristos, and Olivo]{attia2019review}
A.~B.~E. Attia, G.~Balasundaram, M.~Moothanchery, U.~Dinish, R.~Bi,
  V.~Ntziachristos, and M.~Olivo.
\newblock A review of clinical photoacoustic imaging: Current and future
  trends.
\newblock \emph{Photoacoustics}, 16:\penalty0 100144, 2019.

\bibitem[Beard(2011)]{beard2011biomedical}
P.~Beard.
\newblock Biomedical photoacoustic imaging.
\newblock \emph{Interface Focus}, 1\penalty0 (4):\penalty0 602--631, 2011.

\bibitem[Becker et~al.(2018)Becker, Masthoff, Claussen, Ford, Roll, Burg,
  Barth, Heindel, Schaefers, Eisenblaetter, et~al.]{becker2018multispectral}
A.~Becker, M.~Masthoff, J.~Claussen, S.~J. Ford, W.~Roll, M.~Burg, P.~J. Barth,
  W.~Heindel, M.~Schaefers, M.~Eisenblaetter, et~al.
\newblock Multispectral optoacoustic tomography of the human breast:
  characterisation of healthy tissue and malignant lesions using a hybrid
  ultrasound-optoacoustic approach.
\newblock \emph{European Radiology}, 28\penalty0 (2):\penalty0 602--609, 2018.

\bibitem[Brunker et~al.(2017)Brunker, Yao, Laufer, and
  Bohndiek]{brunker2017photoacoustic}
J.~Brunker, J.~Yao, J.~Laufer, and S.~E. Bohndiek.
\newblock Photoacoustic imaging using genetically encoded reporters: a review.
\newblock \emph{Journal of Biomedical Optics}, 22\penalty0 (7):\penalty0
  070901, 2017.

\bibitem[Chlis et~al.(2020)Chlis, Karlas, Fasoula, Kallmayer, Eckstein, Theis,
  Ntziachristos, and Marr]{chlis2020sparse}
N.-K. Chlis, A.~Karlas, N.-A. Fasoula, M.~Kallmayer, H.-H. Eckstein, F.~J.
  Theis, V.~Ntziachristos, and C.~Marr.
\newblock A sparse deep learning approach for automatic segmentation of human
  vasculature in multispectral optoacoustic tomography.
\newblock \emph{Photoacoustics}, 20:\penalty0 100203, 2020.

\bibitem[Dice(1945)]{dice1945measures}
L.~R. Dice.
\newblock Measures of the amount of ecologic association between species.
\newblock \emph{Ecology}, 26\penalty0 (3):\penalty0 297--302, 1945.

\bibitem[Gerl et~al.(2020)Gerl, Paetzold, He, Ezhov, Shit, Kofler, Bayat,
  Tetteh, Ntziachristos, and Menze]{gerl2020distance}
S.~Gerl, J.~C. Paetzold, H.~He, I.~Ezhov, S.~Shit, F.~Kofler, A.~Bayat,
  G.~Tetteh, V.~Ntziachristos, and B.~Menze.
\newblock A distance-based loss for smooth and continuous skin layer
  segmentation in optoacoustic images.
\newblock In \emph{International Conference on Medical Image Computing and
  Computer-Assisted Intervention}, pages 309--319. Springer, 2020.

\bibitem[Gr{\"o}hl et~al.(2021{\natexlab{a}})Gr{\"o}hl, Kirchner, Adler,
  Hacker, Holzwarth, Hern{\'a}ndez-Aguilera, Herrera, Santos, Bohndiek, and
  Maier-Hein]{grohl2021learned}
J.~Gr{\"o}hl, T.~Kirchner, T.~J. Adler, L.~Hacker, N.~Holzwarth,
  A.~Hern{\'a}ndez-Aguilera, M.~A. Herrera, E.~Santos, S.~E. Bohndiek, and
  L.~Maier-Hein.
\newblock Learned spectral decoloring enables photoacoustic oximetry.
\newblock \emph{Scientific Reports}, 11\penalty0 (1):\penalty0 1--12,
  2021{\natexlab{a}}.

\bibitem[Gr{\"o}hl et~al.(2021{\natexlab{b}})Gr{\"o}hl, Schellenberg, Dreher,
  and Maier-Hein]{grohl2021deep}
J.~Gr{\"o}hl, M.~Schellenberg, K.~Dreher, and L.~Maier-Hein.
\newblock Deep learning for biomedical photoacoustic imaging: A review.
\newblock \emph{Photoacoustics}, page 100241, 2021{\natexlab{b}}.

\bibitem[Hauptmann and Cox(2020)]{hauptmann2020deep}
A.~Hauptmann and B.~T. Cox.
\newblock Deep learning in photoacoustic tomography: Current approaches and
  future directions.
\newblock \emph{Journal of Biomedical Optics}, 25\penalty0 (11):\penalty0
  112903, 2020.

\bibitem[Holzwarth et~al.(2021)Holzwarth, Schellenberg, Gr{\"o}hl, Dreher,
  N{\"o}lke, Seitel, Tizabi, M{\"u}ller-Stich, and
  Maier-Hein]{holzwarth2020tattoo}
N.~Holzwarth, M.~Schellenberg, J.~Gr{\"o}hl, K.~Dreher, J.-H. N{\"o}lke,
  A.~Seitel, M.~D. Tizabi, B.~P. M{\"u}ller-Stich, and L.~Maier-Hein.
\newblock Tattoo tomography: Freehand 3d photoacoustic image reconstruction
  with an optical pattern.
\newblock \emph{International Journal of Computer Assisted Radiology and
  Surgery}, pages 1--10, 2021.

\bibitem[Isensee et~al.(2021)Isensee, Jaeger, Kohl, Petersen, and
  Maier-Hein]{isensee2021nnu}
F.~Isensee, P.~F. Jaeger, S.~A. Kohl, J.~Petersen, and K.~H. Maier-Hein.
\newblock nn{U-Net}: a self-configuring method for deep learning-based
  biomedical image segmentation.
\newblock \emph{Nature Methods}, 18\penalty0 (2):\penalty0 203--211, 2021.

\bibitem[Kirchner et~al.(2018)Kirchner, Sattler, Gr{\"o}hl, and
  Maier-Hein]{kirchner2018signed}
T.~Kirchner, F.~Sattler, J.~Gr{\"o}hl, and L.~Maier-Hein.
\newblock Signed real-time delay multiply and sum beamforming for multispectral
  photoacoustic imaging.
\newblock \emph{Journal of Imaging}, 4\penalty0 (10):\penalty0 121, 2018.

\bibitem[Lafci et~al.(2020{\natexlab{a}})Lafci, Mer{\v{c}}ep, Morscher,
  De{\'a}n-Ben, and Razansky]{lafci2020deep}
B.~Lafci, E.~Mer{\v{c}}ep, S.~Morscher, X.~L. De{\'a}n-Ben, and D.~Razansky.
\newblock Deep learning for automatic segmentation of hybrid optoacoustic
  ultrasound (opus) images.
\newblock \emph{IEEE Transactions on Ultrasonics, Ferroelectrics, and Frequency
  Control}, 68\penalty0 (3):\penalty0 688--696, 2020{\natexlab{a}}.

\bibitem[Lafci et~al.(2020{\natexlab{b}})Lafci, Mer{\'c}ep, Morscher,
  De{\'a}n-Ben, and Razansky]{lafci2020efficient}
B.~Lafci, E.~Mer{\'c}ep, S.~Morscher, X.~L. De{\'a}n-Ben, and D.~Razansky.
\newblock Efficient segmentation of multi-modal optoacoustic and ultrasound
  images using convolutional neural networks.
\newblock In \emph{Photons Plus Ultrasound: Imaging and Sensing 2020}, volume
  11240, page 112402N. International Society for Optics and Photonics,
  2020{\natexlab{b}}.

\bibitem[Liang et~al.(2021)Liang, Zhang, Wu, Li, Zhuang, Feng, Chen, and
  Qi]{Liang2021}
Z.~Liang, S.~Zhang, J.~Wu, X.~Li, Z.~Zhuang, Q.~Feng, W.~Chen, and L.~Qi.
\newblock Automatic 3-d segmentation and volumetric light fluence correction
  for photoacoustic tomography based on optimal 3-d graph search.
\newblock \emph{Medical Image Analysis}, page 102275, 2021.
\newblock ISSN 1361-8415.
\newblock \doi{https://doi.org/10.1016/j.media.2021.102275}.

\bibitem[Liu et~al.(2021)Liu, Chen, Zhang, Zhu, and Wang]{liu2021five}
C.~Liu, J.~Chen, Y.~Zhang, J.~Zhu, and L.~Wang.
\newblock Five-wavelength optical-resolution photoacoustic microscopy of blood
  and lymphatic vessels.
\newblock \emph{Advanced Photonics}, 3\penalty0 (1):\penalty0 016002, 2021.

\bibitem[Luke et~al.(2019)Luke, Hoffer-Hawlik, Van~Namen, and
  Shang]{luke2019net}
G.~P. Luke, K.~Hoffer-Hawlik, A.~C. Van~Namen, and R.~Shang.
\newblock O-net: a convolutional neural network for quantitative photoacoustic
  image segmentation and oximetry.
\newblock \emph{arXiv preprint arXiv:1911.01935}, 2019.

\bibitem[Ly et~al.(2021)Ly, Nguyen, Vo, Mondal, Park, Choi, Vu, Kim, and
  Oh]{Ly2021}
C.~D. Ly, V.~T. Nguyen, T.~H. Vo, S.~Mondal, S.~Park, J.~Choi, T.~T.~H. Vu,
  C.-S. Kim, and J.~Oh.
\newblock Full-view in vivo skin and blood vessels profile segmentation in
  photoacoustic imaging based on deep learning.
\newblock \emph{Photoacoustics}, page 100310, 2021.
\newblock ISSN 2213-5979.
\newblock \doi{https://doi.org/10.1016/j.pacs.2021.100310}.

\bibitem[Mongan et~al.(2020)Mongan, Moy, and Kahn]{Mongan2020}
J.~Mongan, L.~Moy, and C.~E. Kahn.
\newblock Checklist for artificial intelligence in medical imaging (claim): A
  guide for authors and reviewers.
\newblock \emph{Radiology: Artificial Intelligence}, 2\penalty0 (2):\penalty0
  e200029, 2020.
\newblock \doi{10.1148/ryai.2020200029}.
\newblock PMID: 33937821.

\bibitem[Moustakidis et~al.(2019)Moustakidis, Omar, Aguirre, Mohajerani, and
  Ntziachristos]{moustakidis2019fully}
S.~Moustakidis, M.~Omar, J.~Aguirre, P.~Mohajerani, and V.~Ntziachristos.
\newblock Fully automated identification of skin morphology in raster-scan
  optoacoustic mesoscopy using artificial intelligence.
\newblock \emph{Medical Physics}, 46\penalty0 (9):\penalty0 4046--4056, 2019.

\bibitem[Nikolov et~al.(2021)Nikolov, Blackwell, Zverovitch, Mendes, Livne,
  De~Fauw, Patel, Meyer, Askham, Romera-Paredes, et~al.]{nikolov}
S.~Nikolov, S.~Blackwell, A.~Zverovitch, R.~Mendes, M.~Livne, J.~De~Fauw,
  Y.~Patel, C.~Meyer, H.~Askham, B.~Romera-Paredes, et~al.
\newblock Clinically applicable segmentation of head and neck anatomy for
  radiotherapy: Deep learning algorithm development and validation study.
\newblock \emph{Journal of Medical Internet Research}, 23\penalty0
  (7):\penalty0 e26151, 2021.

\bibitem[Nolden et~al.(2013)Nolden, Zelzer, Seitel, Wald, M{\"u}ller, Franz,
  Maleike, Fangerau, Baumhauer, Maier-Hein, et~al.]{nolden2013medical}
M.~Nolden, S.~Zelzer, A.~Seitel, D.~Wald, M.~M{\"u}ller, A.~M. Franz,
  D.~Maleike, M.~Fangerau, M.~Baumhauer, L.~Maier-Hein, et~al.
\newblock The medical imaging interaction toolkit: challenges and advances.
\newblock \emph{International journal of computer assisted radiology and
  surgery}, 8\penalty0 (4):\penalty0 607--620, 2013.

\bibitem[Paszke et~al.(2019)Paszke, Gross, Massa, Lerer, Bradbury, Chanan,
  Killeen, Lin, Gimelshein, Antiga, Desmaison, Kopf, Yang, DeVito, Raison,
  Tejani, Chilamkurthy, Steiner, Fang, Bai, and Chintala]{pytorch}
A.~Paszke, S.~Gross, F.~Massa, A.~Lerer, J.~Bradbury, G.~Chanan, T.~Killeen,
  Z.~Lin, N.~Gimelshein, L.~Antiga, A.~Desmaison, A.~Kopf, E.~Yang, Z.~DeVito,
  M.~Raison, A.~Tejani, S.~Chilamkurthy, B.~Steiner, L.~Fang, J.~Bai, and
  S.~Chintala.
\newblock Pytorch: An imperative style, high-performance deep learning library.
\newblock In \emph{Advances in Neural Information Processing Systems 32}, pages
  8024--8035. Curran Associates, Inc., 2019.

\bibitem[Reinke et~al.(2021)Reinke, Eisenmann, Tizabi, Sudre, R{\"a}dsch,
  Antonelli, Arbel, Bakas, Cardoso, Cheplygina, et~al.]{reinke2021common}
A.~Reinke, M.~Eisenmann, M.~D. Tizabi, C.~H. Sudre, T.~R{\"a}dsch,
  M.~Antonelli, T.~Arbel, S.~Bakas, M.~J. Cardoso, V.~Cheplygina, et~al.
\newblock Common limitations of image processing metrics: A picture story.
\newblock \emph{arXiv preprint arXiv:2104.05642}, 2021.

\bibitem[Ren et~al.(2021)Ren, De{\'a}n-Ben, Augath, and
  Razansky]{ren2021feasibility}
W.~Ren, X.~L. De{\'a}n-Ben, M.-A. Augath, and D.~Razansky.
\newblock Feasibility study on concurrent optoacoustic tomography and magnetic
  resonance imaging.
\newblock In \emph{Photons Plus Ultrasound: Imaging and Sensing 2021}, volume
  11642, page 116420C. International Society for Optics and Photonics, 2021.

\bibitem[Ronneberger et~al.(2015)Ronneberger, Fischer, and
  Brox]{ronneberger2015u}
O.~Ronneberger, P.~Fischer, and T.~Brox.
\newblock U-{Net}: Convolutional networks for biomedical image segmentation.
\newblock In \emph{International Conference on Medical Image Computing and
  Computer-Assisted Intervention}, pages 234--241. Springer, 2015.

\bibitem[Ro{\ss} et~al.(2021)Ro{\ss}, Bruno, Reinke, Wiesenfarth, Koeppel,
  Full, Pekdemir, Godau, Trofimova, Isensee, et~al.]{ross2021can}
T.~Ro{\ss}, P.~Bruno, A.~Reinke, M.~Wiesenfarth, L.~Koeppel, P.~M. Full,
  B.~Pekdemir, P.~Godau, D.~Trofimova, F.~Isensee, et~al.
\newblock How can we learn (more) from challenges? {A statistical approach to
  driving future algorithm development}.
\newblock \emph{arXiv preprint arXiv:2106.09302}, 2021.

\bibitem[Tan and Le(2019)]{tan2019efficientnet}
M.~Tan and Q.~Le.
\newblock Efficientnet: Rethinking model scaling for convolutional neural
  networks.
\newblock In \emph{International Conference on Machine Learning}, pages
  6105--6114. PMLR, 2019.

\bibitem[Triki and Xue(2021)]{triki2021h}
F.~Triki and Q.~Xue.
\newblock H\"older stability of quantitative photoacoustic tomography based on
  partial data.
\newblock \emph{Inverse Problems}, 37\penalty0 (10):\penalty0 105007, 2021.
\newblock \doi{10.1088/1361-6420/ac1e7e}.

\bibitem[Wei et~al.(2015)Wei, Nguyen, Xia, Arnal, Wong, Pelivanov, and
  O'Donnell]{wei2015real}
C.-W. Wei, T.-M. Nguyen, J.~Xia, B.~Arnal, E.~Y. Wong, I.~M. Pelivanov, and
  M.~O'Donnell.
\newblock Real-time integrated photoacoustic and ultrasound (paus) imaging
  system to guide interventional procedures: ex vivo study.
\newblock \emph{IEEE Transactions on Ultrasonics, Ferroelectrics, and Frequency
  Control}, 62\penalty0 (2):\penalty0 319--328, 2015.

\bibitem[Wiesenfarth et~al.(2021)Wiesenfarth, Reinke, Landman, Eisenmann, Saiz,
  Cardoso, Maier-Hein, and Kopp-Schneider]{wiesenfarth2021methods}
M.~Wiesenfarth, A.~Reinke, B.~A. Landman, M.~Eisenmann, L.~A. Saiz, M.~J.
  Cardoso, L.~Maier-Hein, and A.~Kopp-Schneider.
\newblock Methods and open-source toolkit for analyzing and visualizing
  challenge results.
\newblock \emph{Scientific Reports}, 11\penalty0 (1):\penalty0 1--15, 2021.

\bibitem[Xia et~al.(2014)Xia, Yao, and Wang]{xia2014photoacoustic}
J.~Xia, J.~Yao, and L.~V. Wang.
\newblock Photoacoustic tomography: principles and advances.
\newblock \emph{Electromagnetic Waves (Cambridge, Mass.)}, 147:\penalty0 1,
  2014.

\bibitem[Yuan et~al.(2020)Yuan, Gao, Peng, Zhou, Liu, Zhu, and
  Song]{yuan2020hybrid}
A.~Y. Yuan, Y.~Gao, L.~Peng, L.~Zhou, J.~Liu, S.~Zhu, and W.~Song.
\newblock Hybrid deep learning network for vascular segmentation in
  photoacoustic imaging.
\newblock \emph{Biomedical Optics Express}, 11\penalty0 (11):\penalty0
  6445--6457, 2020.

\bibitem[Zhang et~al.(2021)Zhang, Bo, Wang, Di~Spirito, Huang, Nyayapathi,
  Zheng, Vu, Gong, Yao, et~al.]{zhang2021deep}
H.~Zhang, W.~Bo, D.~Wang, A.~Di~Spirito, C.~Huang, N.~Nyayapathi, E.~Zheng,
  T.~Vu, Y.~Gong, J.~Yao, et~al.
\newblock Deep-{E}: A fully-dense neural network for improving the elevation
  resolution in linear-array-based photoacoustic tomography.
\newblock \emph{IEEE Transactions on Medical Imaging}, 2021.

\bibitem[Zhang et~al.(2018)Zhang, Chen, Zhou, Lan, and
  Gao]{zhang2018photoacoustic}
J.~Zhang, B.~Chen, M.~Zhou, H.~Lan, and F.~Gao.
\newblock Photoacoustic image classification and segmentation of breast cancer:
  a feasibility study.
\newblock \emph{IEEE Access}, 7:\penalty0 5457--5466, 2018.

\bibitem[Zlateski et~al.(2018)Zlateski, Jaroensri, Sharma, and
  Durand]{zlateski2018importance}
A.~Zlateski, R.~Jaroensri, P.~Sharma, and F.~Durand.
\newblock On the importance of label quality for semantic segmentation.
\newblock In \emph{Proceedings of the IEEE Conference on Computer Vision and
  Pattern Recognition}, pages 1479--1487, 2018.

\end{thebibliography}
\end{document}